\documentclass{article}
\usepackage{log_2023}						

\usepackage{booktabs}	
\usepackage{amssymb}
\usepackage{pifont}
\newcommand{\xmark}{\ding{55}}%
\usepackage{multirow}			
\usepackage{graphicx,wrapfig}
\usepackage{amsfonts}						
\usepackage{graphicx}	

\usepackage{soul}
\usepackage{duckuments}

\usepackage[numbers,compress,sort]{natbib}	

\title{Graph Neural Networks for Recommendation: Reproducibility, Graph Topology, and Node Representation}

\author[Daniele Malitesta and Claudio Pomo and Tommaso Di Noia]{%
Daniele Malitesta\\
Polytechnic University of Bari, Italy\\
\email{daniele.malitesta@poliba.it}\And
Claudio Pomo\\
Polytechnic University of Bari, Italy\\
\email{claudio.pomo@poliba.it}\And
Tommaso Di Noia\\
Polytechnic University of Bari, Italy\\
\email{tommaso.dinoia@poliba.it}
}
\begin{document}

\maketitle

\begin{abstract}
Graph neural networks (GNNs) have gained prominence in recommendation systems in recent years. By representing the user-item matrix as a bipartite and undirected graph, GNNs have demonstrated their potential to capture short- and long-distance user-item interactions, thereby learning more accurate preference patterns than traditional recommendation approaches. In contrast to previous tutorials on the same topic, this tutorial aims to present and examine three key aspects that characterize GNNs for recommendation: (i) the reproducibility of state-of-the-art approaches, (ii) the potential impact of graph topological characteristics on the performance of these models, and (iii) strategies for learning node representations when training features from scratch or utilizing pre-trained embeddings as additional item information (e.g., multimodal features). The goal is to provide three novel theoretical and practical perspectives on the field, currently subject to debate in graph learning but long been overlooked in the context of recommendation systems.
\end{abstract}

\section{Learning objectives}

With the current tutorial, we plan to cover both theoretical and practical aspects in GNNs-based recommendation. First, we investigate the current challenges in experimentally-reproducing the results of state-of-the-art approaches, and compare their performance to other shallower models for recommendation (with unexpected outcomes)~\cite{DBLP:conf/recsys/AnelliMPBSN23, DBLP:conf/um/MalitestaPANF23}. For the experimental study, we adopt Elliot~\cite{DBLP:conf/sigir/AnelliBFMMPDN21}, our framework for the rigorous reproducibility and evaluation of recommender systems. Indeed, such an investigation paves the way to understanding whether, how, and why topological graph properties (conceptually related to node degree) may influence the performance of GNNs-based recommendation systems~\cite{DBLP:journals/corr/abs-2308-10778}. Finally, the tutorial delves into the representation strategies for node embeddings~\cite{DBLP:conf/ecir/AnelliDNMPP23, DBLP:conf/recsys/AnelliDNSFMP22}; while most of the techniques learn node representation from scratch, our focus goes to the adoption of pre-trained item's side information (such as multimodal ones~\cite{DBLP:journals/corr/abs-2309-05273}) by analysing the main experimental implications of such a design choice~\cite{DBLP:journals/corr/abs-2308-12911}. 

\section{Tutorial schedule}

November 30, 2023 (5pm-8pm GMT), Online.

Total tutorial duration $\rightarrow$ \textit{180 minutes}

\vspace{2mm}
\noindent \textbf{\href{https://sisinflab.github.io/tutorial-gnns-recsys-log2023/assets/slides/Part0.pdf}{Introduction and background}} (\ul{Tommaso Di Noia}) $\rightarrow$ \textit{20 minutes}
\begin{itemize}
    \item Introduction and motivations of the tutorial $\rightarrow$ \textit{5 minutes}
    \item Basics concepts of recommender systems \& GNNs-based recommendation $\rightarrow$ \textit{15 minutes}
\end{itemize}

\vspace{2mm}
\noindent \textbf{\href{https://sisinflab.github.io/tutorial-gnns-recsys-log2023/assets/slides/Part1.pdf}{Reproducibility}} (\ul{Claudio Pomo}) $\rightarrow$ \textit{60 minutes}
\begin{itemize}
    \item \textbf{[\href{https://colab.research.google.com/drive/1_li7RQ_Rj4JaAVpw1kvuOGhrDpfCL-UQ?usp=sharing}{Hands-on \#1}]} Implementation and reproducibility of GNNs-based recsys in Elliot with PyG and reproducibility issues $\rightarrow$ \textit{35 minutes}
    \item Performance comparison of GNNs-based approaches to traditional recommendation systems $\rightarrow$ \textit{25 minutes}
\end{itemize}

\vspace{2mm}
\noindent \textbf{Break and Q\&A} $\rightarrow$ \textit{15 minutes} 

\vspace{2mm}
\noindent \textbf{\href{https://sisinflab.github.io/tutorial-gnns-recsys-log2023/assets/slides/Part2.pdf}{Graph topology}} $\rightarrow$ \textit{30 minutes}
\begin{itemize}
    \item Concepts and formulations of graph topological properties of the user-item graph (\ul{Tommaso Di Noia}) $\rightarrow$ \textit{15 minutes}
    \item Impact of topological graph properties on the performance of GNNs-based recommender systems (\ul{Daniele Malitesta}) $\rightarrow$ \textit{15 minutes}
\end{itemize}

\vspace{2mm}
\noindent \textbf{\href{https://sisinflab.github.io/tutorial-gnns-recsys-log2023/assets/slides/Part3.pdf}{Node representation}} (\ul{Daniele Malitesta}) $\rightarrow$ \textit{45 minutes}
\begin{itemize}
    \item Design choices to train node embeddings from scratch $\rightarrow$ \textit{20 minutes}
    \item \textbf{[\href{https://colab.research.google.com/drive/1socyjwzmYNAm3trlquAevq-R1d4zX3KH?usp=sharing}{Hands-on \#2}]} Leveraging item's side-information (e.g., multimodal features) to represent node embeddings $\rightarrow$ \textit{25 minutes}
\end{itemize}

\vspace{2mm}
\noindent \textbf{Closing remarks and Q\&A} $\rightarrow$ \textit{10 minutes}

\section{Relevance to LoG}

The tutorial covers highly-related topics to the ones of the LoG conference and community, namely, the application of graph neural networks (GNNs) to the task of personalized recommendation. In particular, and differently from previous similar tutorials presented at other venues (see later), the main focus of this tutorial is on the reproducibility of state-of-the-art recommendation systems leveraging GNNs, the possible impact of graph topological characteristics on the performance of such approaches, and the modeling of node features (trained from scratch or freezed to use additional side information). Since the outlined three aspects have been widely debated in the graph representation learning literature over the last few years, we believe the LoG conference could benefit from an analysis of these aspects also in the field of recommender systems. 

\section{Previous related tutorials}

\noindent \textbf{Tutorials on graph-based recommendation.} Previous tutorials on graph-based recommendation~\cite{DBLP:conf/wsdm/Wang0C20, DBLP:conf/kdd/El-KishkyBXH22, DBLP:conf/wsdm/GaoW0022, DBLP:conf/um/PurificatoBL23} are reported in Table~\ref{tab:tutorials}, along with the reference, website, slides, and video recording. 

\noindent \textbf{Differences with previous tutorials.} The majority of previous tutorials on graph-based recommendation address the topic of GNNs in recommendation from a general perspective. Conversely, our tutorial intends to approach the same topic from three main research aspects which are highly popular in the graph learning field but have not been previously investigated in graph-based recommendation, namely: reproducibility issues, the influence of graph topology on model's performance, and the modeling of node features. Indeed, the tutorials~\cite{DBLP:conf/um/PurificatoBL23, DBLP:conf/cikm/PurificatoBL23} are the closest to ours in the intention of providing a more specific analysis of GNNs-based recommendation, by considering user modeling and beyond-accuracy evaluation; however, as already highlighted, our investigation and contributions are different.

\noindent \textbf{Other related tutorials.} Scalable Graph Neural Networks with Deep Graph Library (\ul{KDD 2020});  Deep Graph Learning: Foundations, Advances and Applications (\ul{KDD 2020}); Learning Graph Neural Networks with Deep Graph Library (\ul{The Web Conf 2020}); Graph Representation Learning: Foundations, Methods, Applications and Systems (\ul{KDD 2021}); Advanced Deep Graph Learning: Deeper, Faster, Robuster, Unsupervised (\ul{The Web Conf 2021}); Learning from Graphs: From Mathematical Principles to Practical Tools (\ul{The Web Conf 2021}); Scalable Graph Neural Networks with Deep Graph Library (\ul{WSDM 2021}); Frontiers of Graph Neural Networks with DIG (\ul{KDD 2022}); Graph Neural Networks: Foundation, Frontiers and Applications (\ul{KDD 2022}); Graph Neural Networks: Foundation, Frontiers and Applications (\ul{The Web Conf 2023}); Self-supervised Learning and Pre-training on Graphs (\ul{The Web Conf 2023}); When Sparse Meets Dense: Learning Advanced Graph Neural Networks with DGL-Sparse Package (\ul{The Web Conf 2023}).

\begin{table}[!t]
    \caption{Tutorials about graph-based recommendation from 2020-2023.}\label{tab:tutorials}
    \centering
    \begin{tabular}{lcccc}
    \toprule
        \textbf{Title} & \textbf{Venue} & \textbf{Website} & \textbf{Slides} & \textbf{Video} \\ \cmidrule{1-5}
        \parbox[t]{7cm}{Learning and Reasoning on Graph for Recommendation~\cite{DBLP:conf/wsdm/Wang0C20}} & WSDM 2020 & \href{https://next-nus.github.io}{link} & \href{https://next-nus.github.io/slides/tuto-cikm2019-public.pdf}{link} & \xmark \\ \cmidrule{1-5}
        \parbox[t]{7cm}{Graph-based Representation Learning for Web-scale Recommender Systems~\cite{DBLP:conf/kdd/El-KishkyBXH22}} & KDD 2022 & \href{https://ahelk.github.io/talks/ecmlpkdd22.html}{link} & \href{https://ahelk.github.io/talks/kdd22/ecml.pdf}{link} & \xmark \\ \cmidrule{1-5}
         \parbox[t]{7cm}{Graph Neural Networks for Recommender System~\cite{DBLP:conf/wsdm/GaoW0022}} & WSDM 2022 & \href{https://sites.google.com/view/gnn-recsys}{link} & \href{https://drive.google.com/file/d/1VkP0G_M7Kg2492wjxZ_F00DoTC0QyJlB/view?usp=sharing}{link} & \href{https://drive.google.com/file/d/1OQO6JldJC_XyRAxGM59CLyzMmUhIadyq/view?usp=sharing}{link} \\
         \cmidrule{1-5}
         \parbox[t]{7cm}{Tutorial on User Profiling with Graph Neural Networks and Related Beyond-Accuracy Perspectives~\cite{DBLP:conf/um/PurificatoBL23}} & UMAP 2023 & \href{https://beyondaccuracy-userprofiling.github.io/tutorial-umap23/}{link} & \href{https://www.slideshare.net/ErasmoPurificato2/tutorial-on-user-profiling-with-graph-neural-networks-and-related-beyondaccuracy-perspectives}{link} & \xmark \\ \cmidrule{1-5}
         \parbox[t]{7cm}{Leveraging Graph Neural Networks for User Profiling: Recent Advances and Open Challenges~\cite{DBLP:conf/cikm/PurificatoBL23}} & CIKM 2023 & \href{https://beyondaccuracy-userprofiling.github.io/tutorial-cikm23/}{link} & \href{https://www.slideshare.net/ErasmoPurificato2/leveraging-graph-neural-networks-for-user-profiling-recent-advances-and-open-challenges}{link} & \xmark\\
        \bottomrule
    \end{tabular}
\end{table}

\section{Useful materials}
All useful materials are available at the tutorial's website:~\url{https://sisinflab.github.io/tutorial-gnns-recsys-log2023/}, along with the GitHub repository:~\url{https://github.com/sisinflab/LoG-2023-GNNs-RecSys}.

\section{Tutorial speakers}

\begin{wrapfigure}{l}{0.16\textwidth}
  \includegraphics[width=\linewidth]{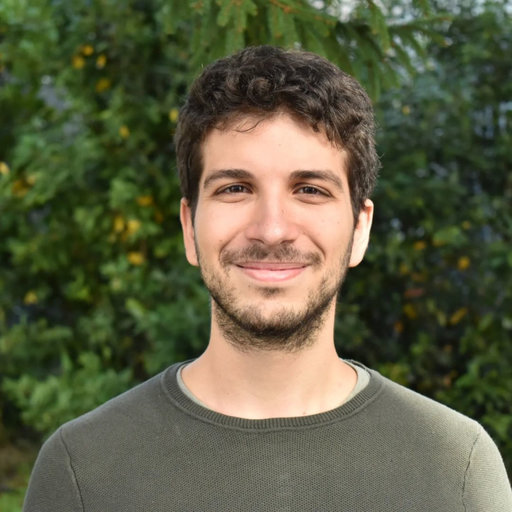}
\end{wrapfigure}
\noindent \textbf{Daniele Malitesta} is a PhD candidate at the Polytechnic University of Bari (Italy). During his research career so far, he has been studying and developing recommendation algorithms leveraging side information, with a specific focus on graph- and multimodal-based recommender systems. He has published at top-tier conferences, such as SIGIR, ECIR, RecSys, and MM, and has served as a reviewer at SIGIR 2023, RecSys 2023 (outstanding reviewer), NeurIPS 2023, LoG 2022-2023, ICLR 2024, and ECIR 2024. He is one of the organizers of the First International Workshop on Graph-Based Approaches in Information Retrieval (IRonGraphs), co-located with ECIR 2024. He is among the developers of Elliot, a framework for the rigorous evaluation and reproducibility of recommender systems, where he contributed with the implementation of more than 15 algorithms from the state-of-the-art. Recently, he has visited Dr. Pasquale Minervini at the University of Edinburgh as part of the internship period of his PhD.

\vspace{2mm}

\begin{wrapfigure}{l}{0.16\textwidth}
  \includegraphics[width=\linewidth]{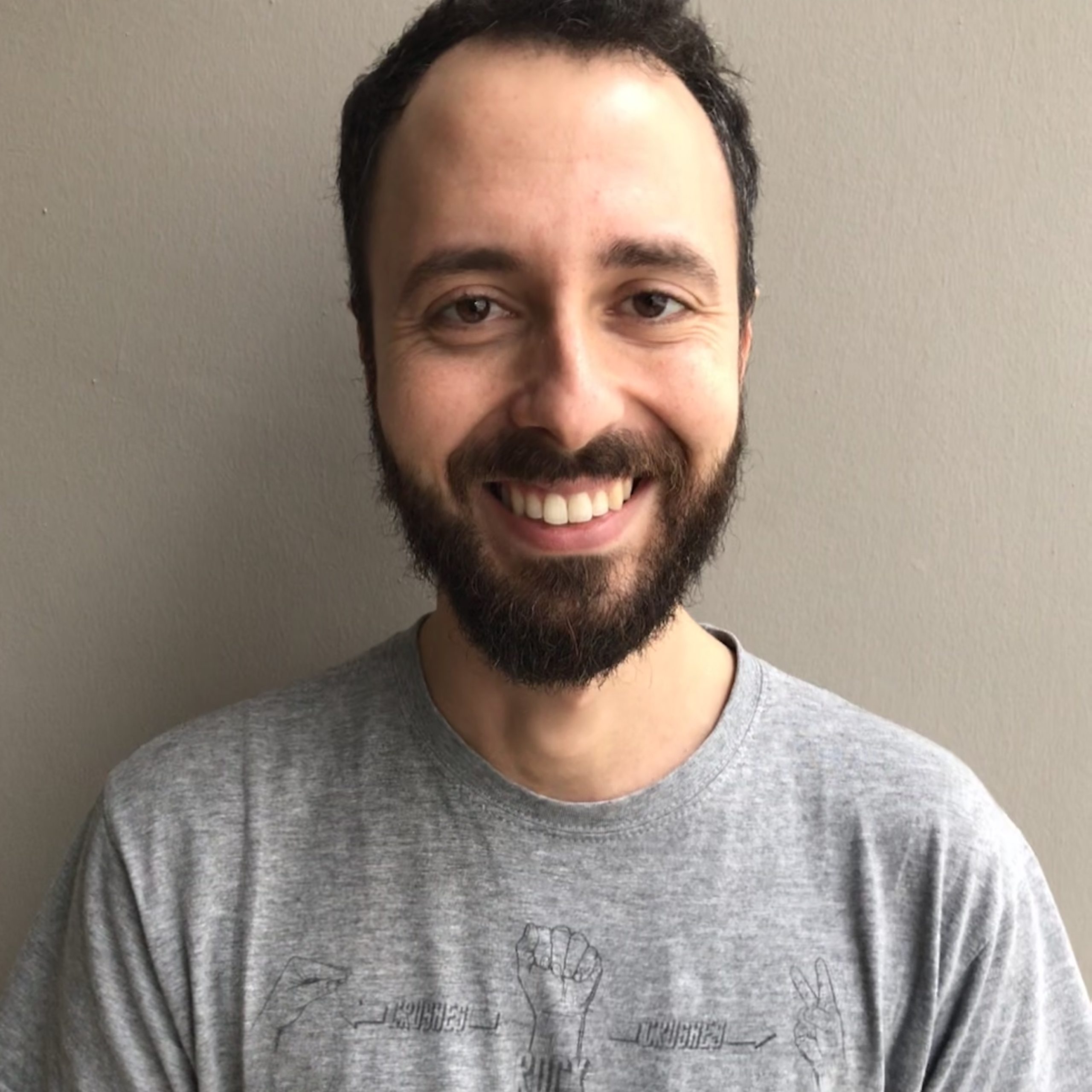}
\end{wrapfigure}
\noindent \textbf{Claudio Pomo} is a research fellow at the Polytechnic University of Bari in Italy, where he obtained his doctorate in computer engineering. His research focuses on responsible AI for personalization, with a particular emphasis on reproducibility of results and multi-objective performance evaluation. Claudio has made significant contributions in these areas, with his work being accepted at prominent conferences such as SIGIR, RecSys, ECIR, UMAP, and in journals including Information Science and IPM. He has also actively participated in the academic community, serving as a reviewer for conferences like SIGIR, RecSys, NeurIPS, WSDM, ECIR, and UMAP. Claudio delivered a tutorial at RecSys21 titled "Pursuing Privacy in Recommender Systems: the View of Users and Researchers from Regulations to Applications." More recently, he co-organized a workshop at KDD 2023 focused on recommender system evaluation, known as EvalRS. Claudio is also one of the authors and contributors to Elliot, a framework designed to assess the rigorous evaluation and reproducibility of recommender systems.

\vspace{2mm}

\begin{wrapfigure}{l}{0.16\textwidth}
  \includegraphics[width=\linewidth]{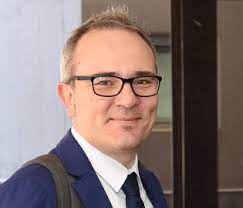}
\end{wrapfigure}
\noindent \textbf{Tommaso Di Noia} is Professor of Computer Science at the Polytechnic University of Bari
(Italy). His research activities focus on AI and Data Management. They were initially devoted to knowledge representation and automated reasoning. Then, he studied how to apply knowledge representation techniques to automated negotiations. Following these ideas, he has devoted his interest to applying knowledge graphs and Linked Open Data to RSs with papers published in international journals, conferences, and book chapters. During the last years, he moved his research into the Trustworthy AI topic with a particular interest in adversarial ML, explainability, fairness and privacy protection of RSs. He is serving as program co-chair at RecSys 2023 and general co-chair at RecSys 2024.

\section{Intended audience and level}

The proposed tutorial deals with both intermediate and advanced theoretical/practical topics spanning recommendation, graph representation learning, reproducibility in machine learning, graph topology, and multimodal learning. For these reasons, the tutorial might ideally be of interest to a wide range of researchers and practitioners working on (even a subset of) such aspects. Programming knowledge of Python and PyTorch (with a specific focus on PyTorch Geometric) would be a good-to-have skill, even though the tutorial will guide the attendees step-by-step, especially during the hands-on sessions. Given the virtual nature of the tutorial, we do not set a strong requirement for the maximum number of participants (ideally up to 200). 

\bibliographystyle{unsrtnat}
\bibliography{reference}

\end{document}